\documentclass[11pt]{article}
\usepackage{booktabs} % For professional-looking tables
\usepackage{xcolor}
\usepackage{graphicx}
\usepackage{array}
\usepackage{setspace}
\usepackage{subcaption} 
\usepackage{listings}
\usepackage{caption}
\usepackage{float}
\usepackage{authblk}

\renewcommand{\arraystretch}{1.5} 
\usepackage{hyperref}
\definecolor{mycolor}{RGB}{0,88,204}
\hypersetup{
  colorlinks=true,
  linkcolor=mycolor,
  urlcolor=mycolor,
  citecolor=mycolor
}
\usepackage{amsmath}
\usepackage{geometry}

\usepackage{fancyhdr}

\fancypagestyle{plain}{%
  \fancyhf{}  
  \fancyhead[R]{\thepage}  
    
}
\usepackage{mathptmx}

\usepackage{titlesec}

\titleformat{\section}
  {\normalfont\large\centering}{\thesection.}{1em}{} 
\titleformat{\subsection}
  {\normalfont\centering}{\thesubsection.}{1em}{} 
  
  \titlespacing{\paragraph}{10pt}{0pt}{6pt}[0pt]
\usepackage{listings}
\usepackage[T1]{fontenc}
\usepackage[utf8]{inputenc}
\usepackage{amssymb}

\definecolor{keywordcolor}{rgb}{0.7, 0.1, 0.1}   % red
\definecolor{tacticcolor}{rgb}{0.0, 0.1, 0.6}    % blue
\definecolor{commentcolor}{rgb}{0.4, 0.4, 0.4}   % grey
\definecolor{symbolcolor}{rgb}{0.0, 0.1, 0.6}    % blue
\definecolor{sortcolor}{rgb}{0.1, 0.5, 0.1}      % green
\definecolor{attributecolor}{rgb}{0.7, 0.1, 0.1} % red

\lstnewenvironment{code}[1][]%
{
   \noindent\newline
   \minipage{\linewidth} 
   \vspace{0.5\baselineskip}
   \lstset{
 	frame = lrtb, 
 	rulecolor=\color{mycolor},
 	escapeinside={/*!}{!*/},
	language=lean, 
	aboveskip = 5mm,
	belowskip = 5mm,
	xleftmargin=2mm,
	xrightmargin=2mm,
	}
	}
{\endminipage\newline}
\lstnewenvironment{codeLong}[1][]%
{
   \lstset{
 	frame = lrtb, 
 	rulecolor=\color{mycolor},
 	escapeinside={/*!}{!*/},
	language=lean, 
	aboveskip = 5mm,
	belowskip = 5mm,
	xleftmargin=2mm,
	xrightmargin=2mm,
	}
	}
{}

\title{Benchmarking Energy Calculations Using Formal Proofs}
\author[1]{\normalsize Ejike D. Ugwuanyi}
\author[1]{\normalsize Colin T. Jones}
\author[1]{\normalsize John Velkey}
\author[1,2,*]{\normalsize  Tyler R. Josephson}

\affil[1]{Department of Chemical, Biochemical, and Environmental Engineering, \authorcr University of Maryland, Baltimore County, Baltimore, MD, USA}
\affil[2]{Department of Computer Science and Electrical Engineering, \authorcr University of Maryland, Baltimore County, Baltimore, MD, USA}
\affil[*]{Corresponding author, email: tjo@umbc.edu}

\date{May 15, 2025}
\begin{document}

\maketitle 
%\vspace{-1cm}
\begin{abstract}
\noindent 
Traditional approaches for validating molecular simulations rely on making software open source and transparent, incorporating unit testing, and generally employing human oversight.
We propose an approach that eliminates software errors using formal logic, providing proofs of correctness. We use the Lean theorem prover and programming language to create a rigorous, mathematically verified framework for computing molecular interaction energies.
We demonstrate this in LeanLJ, a package of functions, proofs, and code execution software that implements Lennard-Jones energy calculations in periodic boundaries.
We introduce a strategy that uses polymorphic functions and type classes to bridge formal proofs (about idealized Real numbers) and executable programs (over floating point numbers).
Execution of LeanLJ matches the current gold standard NIST benchmarks, while providing even stronger guarantees, given LeanLJ's grounding in formal mathematics. 
This approach can be extended to formally verified molecular simulations in particular and formally verified scientific computing software, in general. 

\end{abstract}
\\
\section{Introduction}
Molecular simulations constitute an essential computational tool for understanding atomic-scale phenomena, playing a critical role in predicting the physicochemical properties underlying fields such as materials science, chemistry, and biophysics \cite{filipe_molecular_2022}. For instance, accurate simulations enable the study of molecular interactions that dictate gas adsorption behaviours in porous materials \cite{sun_accurate_2017} \cite{sun_progress_2015}, influence solvation dynamics in chemical solutions \cite{mark_decomposition_1994}, and affect catalytic reactions on material surfaces \cite{s_alvim_material_2023}. Accurate modelling of particle interactions lies at the core of molecular simulations \cite{leach_molecular_2001} . Among the most fundamental and extensively used models are the Lennard-Jones potential, describing van der Waals interactions, and Coulomb potentials, capturing electrostatic forces. The Lennard-Jones potential, specifically, finds significant application in representing non-bonded interactions in chemically relevant systems such as simple fluids \cite{allen_computer_2017}, noble gas clusters \cite{verlet_computer_1967}, hydrocarbon fluids \cite{horton_transferable_2023}, and molecular adsorption phenomena in zeolites or metal-organic frameworks (MOFs) \cite{zhang_brief_2010} \cite{wang_integrated_2024}. It also describes the interaction between a pair of neutral atoms or molecules based on their distance \cite{schwerdtfeger_lennard_2021} \cite{wang_lennard-jones_2020} \cite{rahman_correlations_1964}, and efficiently captures the balance between attractive and repulsive forces \cite{sun_new_2023} \cite{nicolas_equation_1979}. Coulomb interactions are particularly critical for describing electrolyte solutions \cite{fawcett_electrolyte_2004}, protein-ligand binding affinities \cite{l_wen_charged_2003}, and charged colloidal systems \cite{dean_electrostatics_2014} \cite{l_wen_charged_2003}. \\
\\
\noindent To accurately approximate infinite molecular systems from computationally manageable finite-sized simulations, periodic boundary conditions (PBC) are conventionally employed \cite{yang_comparative_2006} \cite{mizzi_implementation_2021}. In chemical simulations, PBC enables the study of bulk-phase properties without boundary effects, such as predicting phase behaviours in liquid water \cite{palos_current_2024}, ionic solutions \cite{lawson_atomistic_2017}, or polymer melts \cite{andrews_workflow_2022}. Furthermore, the minimum image convention ensures that computational resources are efficiently utilized by considering only the nearest periodic images in calculations \cite{wood_monte_1957}, which is particularly important in simulations of dense chemical environments like ionic liquids or liquid crystal phases.
\\

\noindent Software tools like LAMMPS and Gromacs \cite{plimpton_fast_1995}\cite{abraham_gromacs_2015}\cite{van_der_spoel_gromacs_2005} allow users to simulate the dynamics of large molecular systems. However, the sheer complexity of these software packages and the systems they intend to model presents challenges in making simulations transparent, reproducible, useable by others and extensible (TRUE) \cite{thompson_towards_2020}.
For example, the SAMPL Challenges (Statistical Assessment of the Modelling of Proteins and Ligands) \cite{nicholls_predicting_2008} and the Industrial Fluid Properties Simulation Challenges \cite{friend_establishing_2004}\cite{case_first_2004} task computational researchers to predict the solvation or binding free energies of small molecules or the thermophysical properties of fluids. 
Each year, researchers submit highly variable answers, reflecting differences in modelling choices by the researchers (e.g. force fields, simulation conditions, free-energy extrapolation strategies, etc.), as well as more hidden, subtle differences amongst software packages (e.g. default settings for managing Lennard-Jones cut-off and settings for Ewald summation). 
Projects such as the Molecular Simulation Design Framework (MoSDeF) \cite{craven_achieving_2025} \cite{thompson_towards_2020}\cite{crawford_mosdef-gomc_2023} and the Molecular Sciences Software Institute (MolSSI) \cite{nash_molssi_2022} address these issues by providing reproducible workflows for molecular simulation setup, and by teaching and promoting best practices in software development \cite{Thompson_Molecular_reproducibility}.
Simulation software can also be validated by comparing to benchmarks, such as those on the National Institute of Standards and Technology (NIST) Standard Reference Simulation Website (SRSW) \cite{shen_nist_2017}. \\

\begin{table}[h]
    \centering
    \renewcommand{\arraystretch}{1} % Adjust row spacing
    \setlength{\tabcolsep}{5pt} % Adjust column spacing
    \resizebox{\textwidth}{!}{ % Ensures the table fits within margins
    \begin{tabular}{>{\raggedright\arraybackslash}p{3cm}
                    >{\raggedright\arraybackslash}p{4.5cm}
                    >{\raggedright\arraybackslash}p{5.5cm}
                    >{\raggedright\arraybackslash}p{2cm}}
        \toprule
        \textbf{Category of Error} & \textbf{Example} & \textbf{Intervention} & \textbf{Lean} \\
        \midrule
        Syntax & Not closing parentheses & Editor & Editor \\
        Runtime & Accessing element in list that doesn’t exist & Run program, program gives error message & Editor \\
        Semantic & Missing a minus sign, transposing tensor indices & Human inspection of the code; test-driven development; observing anomalous behaviour & Editor \\
        Floating-point/ Round-off & Subtracting small values from large values, ill-conditioned matrices & Modifying simulation methods, using double precision floats & --\\
        \bottomrule
    \end{tabular}
    }
    \caption{Errors in scientific computing software, and typical interventions. Our goal is to develop an approach to address syntax, runtime, and semantic errors in Lean at the ``editor'' stage, before code is compiled.}
    \label{tab:errors}
\end{table}

\noindent We propose an alternative paradigm for improving reliability of molecular simulations. 
To illustrate, consider the taxonomy of programming errors in Table~\ref{tab:errors}.
The simplest are syntax errors: these are addressed immediately because the code cannot compile, the editor highlights the mistake, and the programmer fixes it.
Runtime errors occur during code execution, and may arise when users run the program under conditions not anticipated by the software developers.
Nonetheless, runtime errors typically provide a helpful error message pointing toward the source of the issue.
The deepest issues are semantic errors in the \emph{meaning} of the software: Python would not complain about misinterpreting a scientific principle or incorrectly transcribing math into code -- it is simply not designed for that.
Floating-point and round-off errors create numerical inaccuracies, since computers do not operate with infinite mathematical precision. These are addressed by judicious choices of simulation settings and algorithm choices, and by checking conditions like energy conservation after simulation completion \cite{merz_testing_2018}. \\

\begin{figure}
    \centering
    \includegraphics[width=0.95\textwidth]{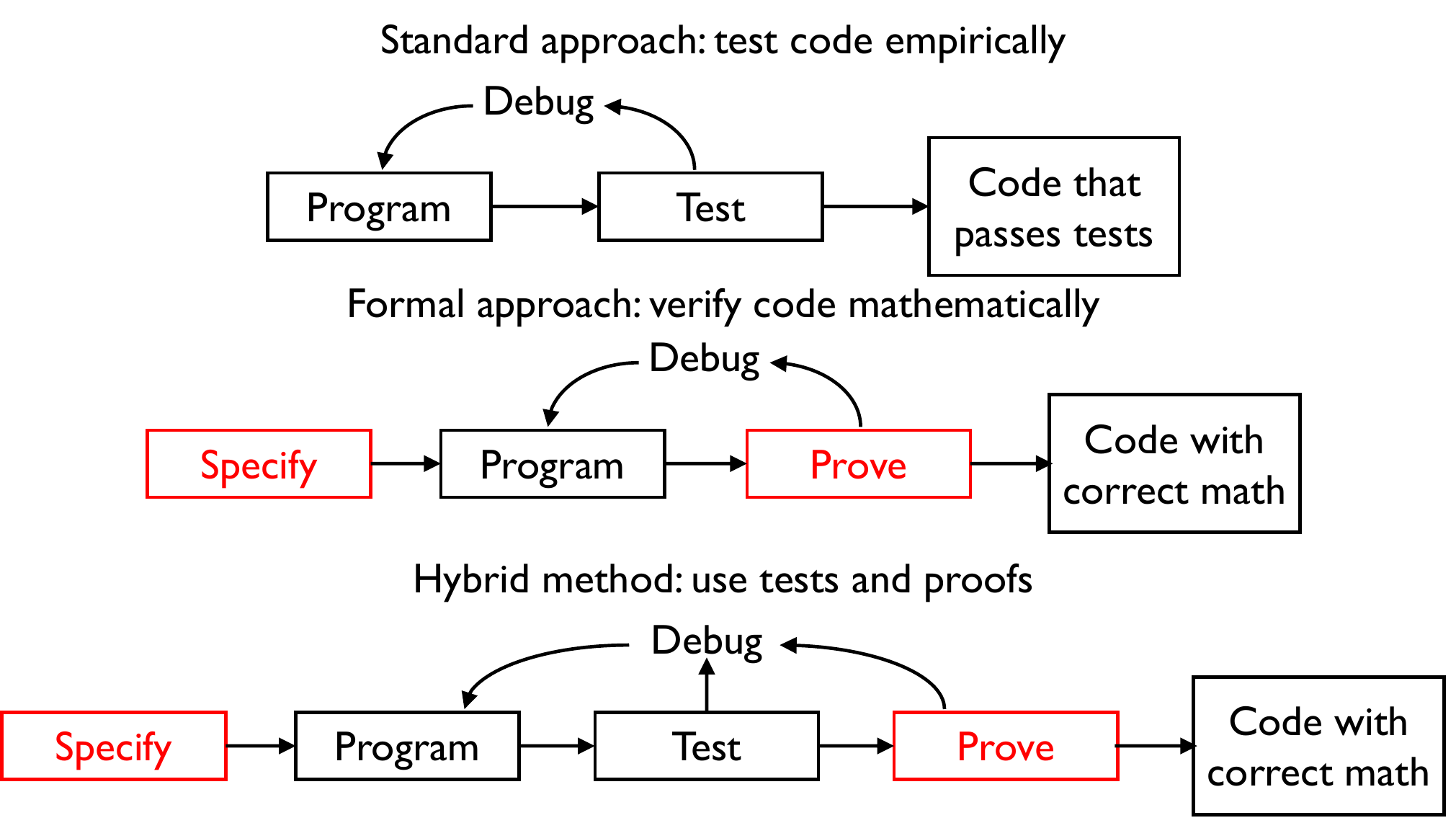}
    \caption{Comparison of code correctness approaches (adapted from \cite{selsam_developing_2017}): the standard test-debug cycle, formal verification using proofs, and a hybrid method combining tests and formal proofs, that we adopt here.}
    \label{fig:correctness-approaches}
\end{figure}

\noindent In this work, we propose a strategy for catching syntax, runtime, and semantic errors at the ``editor'' stage, namely, before the code is compiled.
Our approach stems from the \emph{formal methods} community in computer science, which seeks to prove when software is correct by construction before it is run (also known as static program analysis), unlike traditional testing, which checks for errors by running a program with different inputs.
This approach is handy in areas where even small errors can have significant consequences, such as hardware design and critical software systems.
A prominent example is the Pentium FDIV bug in Intel processors in the early 1990s, the subject of a multi-million dollar recall stemming from a few misplaced bits in chip software \cite{price_pentium_1995}. Now, formal verification approaches prove the correctness of such arithmetic operations in manufactured chips \cite{kaivola_formal_2002}. 
Our approach most closely resembles that of Selsam, et al., who explored how formal methods can be applied to machine learning systems in Certigrad 
\cite{selsam_developing_2017} (Figure \ref{fig:correctness-approaches}).
\noindent By proving the correctness of each step mathematically, this approach exposes errors that might otherwise slip through traditional empirical testing.
They highlight the ability of theorem provers like Lean to eliminate entire classes of high-level errors that arise in complex software systems by enforcing correctness through formal reasoning. They demonstrate their approach by building a variational auto-encoder in Lean, proving properties about their implementation of stochastic gradient descent. \\

\noindent Most prior work on formal methods has focused on floating point operations \cite{bernardo_floating-point_2006}. 
In molecular simulations, these are typically insignificant, but they can lead to issues in certain settings, such as when programs are run with less precision to increase speed, or under extreme conditions.
Tran and Wang \cite{tran_reliable_2017}, explored using interval arithmetic to model the propagation of these uncertainties in molecular dynamics simulations.
Our work sets aside the imprecision of floating-point arithmetic, and instead focuses on verifying higher-level logic and mathematics.
Incorporating interval arithmetic into our approach would in principle be possible, but these tools are currently in development \cite{irving_conservative_2025}. \\

\noindent Lean 4 is a theorem prover and functional programming language designed to write and verify mathematical proofs, as well as write formally-verified software \cite{de_moura_lean_2021}. Unlike traditional programming languages used for scientific computing (C, FORTRAN, Python, etc.), Lean provides a formally verified framework in which proofs of correctness can be explicitly constructed and checked \cite{Christiansen_functional_programming_lean}.
We previously used Lean to formalize chemical physics \cite{bobbin_formalizing_2024}. Lean is also being used to formalize theories in high-energy physics \cite{tooby_physics}. We also recognize Tomáš Skřivan's ongoing SciLean project, which is working out methods for automated differentiation and efficient, array-based computations in Lean \cite{skrivan_lecopivoscilean_2025}. \\

\noindent In our work \cite{bobbin_formalizing_2024}, we showed how theories in science can be rigorously encoded using the Lean theorem prover, proving the correctness of the derivations, grounding them in the foundations of mathematics. We formalized derivations of the Langmuir and BET adsorption models, meticulously defining assumptions and derivations to ensure mathematical rigour. That work was limited to \emph{proofs} in Lean -- we extend that now to executable \emph{programs} with formally-verified properties. \\

\noindent In this paper, we first present a familiar, informal description of Lennard-Jones energy calculations of periodic fluids (Section \ref{sec:informal}). 
We then highlight the proof components (especially definitions and theorems) for the formal implementation in Lean (Section \ref{sec:formal}).
Section \ref{sec:execution} describes how we implement these energy calculations in Lean, which requires novel approaches using functional programming, type polymorphism, and monads.
Section \ref{sec:results} compares our calculations with the results from the NIST SRSW benchmarks \cite{shen_nist_2017}. \\

%%%%
\section{Methods}
\label{sec:Meth}

We implemented this using Lean version 4.16.0-rc2, Mathlib 4 at commit \texttt{e1a3d4c}, and Visual Studio Code version 1.96.
The source code is available in \href{https://github.com/ATOMSLab/LeanLJ}{LeanLJ Repository}.
\\

\section{Informal Description of the Molecular Simulation System}
\label{sec:informal}
The Lennard-Jones system is modelled as a collection of \(N\) particles confined within a cubic simulation box of side length \(L\). The position of each particle is represented as a vector in a three-dimensional space, \(\mathbf{r}_i = (x_i, y_i, z_i)\), where \(i = 1, 2, \dots, N\). The interaction between particles is governed by the Lennard-Jones potential:
\begin{equation}
V_{\text{LJ}}(r_{ij}) = 4\varepsilon \left[ \left( \frac{\sigma}{r_{ij}} \right)^{12} - \left( \frac{\sigma}{r_{ij}} \right)^6 \right]
\label{eq:LJ}
\end{equation}
where \(r_{ij}\) is the distance between particles \(i\) and \(j\), \(\varepsilon\) represents the depth of the potential well, and \(\sigma\) is the characteristic length scale. \\

\begin{figure}[h!]
    \centering
    \begin{subfigure}[t]{0.48\textwidth}
        \centering
        \includegraphics[width=\textwidth]{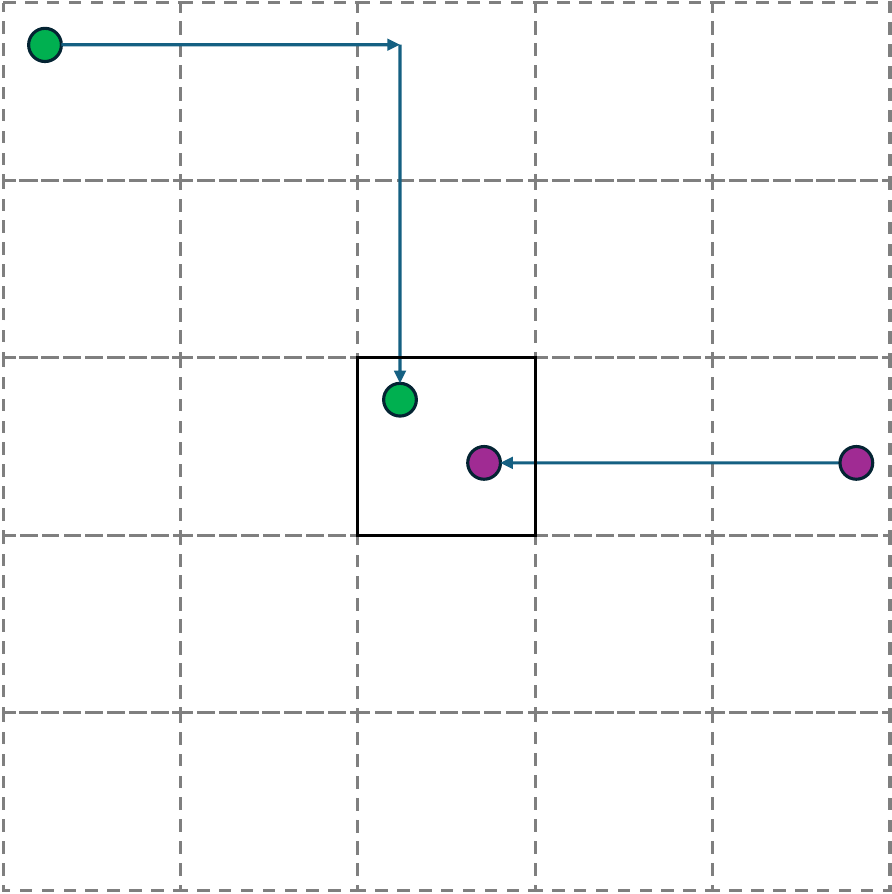}
        \caption{ Periodic boundary conditions.}
    \end{subfigure}
    \hfill
    \begin{subfigure}[t]{0.48\textwidth}
        \centering
        \includegraphics[width=\textwidth]{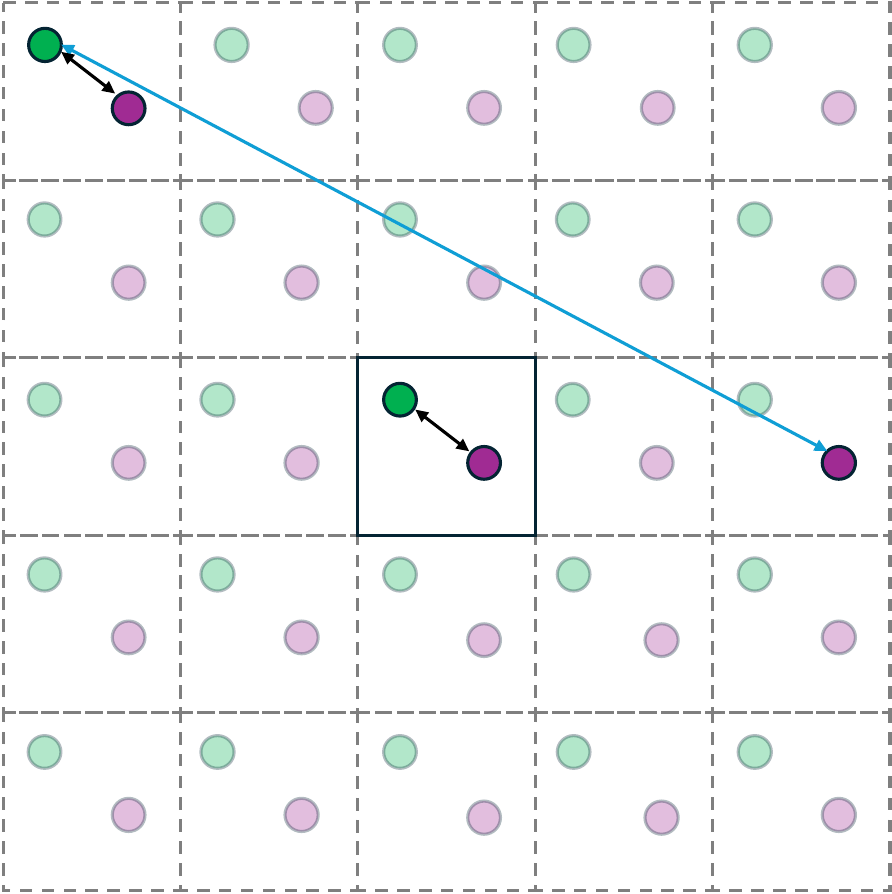}
        \caption{Minimum image convention.}
    \end{subfigure}
    \caption{(a) Periodic boundary conditions: Particles outside the central cubic simulation box are wrapped back into it. Arrows represent the wrapping process along the directions. 
    (b) Minimum image convention: Particles interact with the nearest periodic image, ensuring the shortest distance is used in calculations. The Euclidean distance (blue) is not used; the minimum image distance (black) is used instead, which is equivalent to the minimum image distance between wrapped particles.}
    \label{fig:pbc-mid}
\end{figure}

\noindent Periodic boundary conditions (PBCs) are applied to simulate an infinite system as shown in the equation for particle coordinates in the x, y, and z axes, respectively (Figure \ref{fig:pbc-mid}).

\begin{equation}
x_{i\_\text{wrapped}} = x_i - L \cdot \text{round}\left( \frac{x_i}{L} \right)
\label{eq:pbc_x}
\end{equation}
\begin{equation}
y_{i\_\text{wrapped}} = y_i - L \cdot \text{round}\left( \frac{y_i}{L} \right)
\label{eq:pbc_y}
\end{equation}
\begin{equation}
z_{i\_\text{wrapped}} = z_i - L \cdot \text{round}\left( \frac{z_i}{L} \right)
\label{eq:pbc_z}
\end{equation}

\noindent Because the LJ particles are in a system with PBCs, the distance between two particles is not the Euclidean distance, but the minimum image distance, the shortest pairwise distance considering the periodicity of the box as given in the equation below. \\
\begin{equation}
r_{ij} = \sqrt{
\left( \Delta x - L \cdot \mathrm{round}\left( \frac{\Delta x}{L} \right) \right)^2 +
\left( \Delta y - L \cdot \mathrm{round}\left( \frac{\Delta y}{L} \right) \right)^2 +
\left( \Delta z - L \cdot \mathrm{round}\left( \frac{\Delta z}{L} \right) \right)^2
}
\label{eq:min-image-round}
\end{equation}

\begin{figure}[h!]
    \centering
    \includegraphics[width=0.8\textwidth]{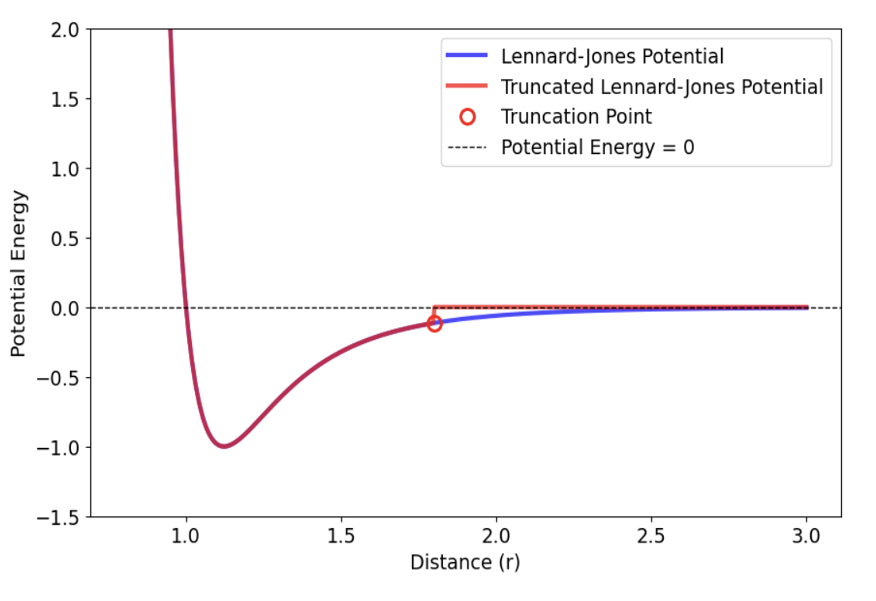} % Reduced width to 80% of text width
    \caption{The Lennard-Jones potential, truncated at the cut-off.}
    \label{fig:lj-potential}
\end{figure}

\noindent To improve computational efficiency, a cut-off radius \(r_{\text{c}}\) is introduced. Interactions are considered only for particle pairs that satisfy \(r_{ij} \leq r_{\text{c}}\), with contributions beyond this radius set to zero. This truncation neglects a relatively minor contribution to the potential energy, depending on the cut-off radius $r_c$ as shown in Figure~\ref{fig:lj-potential}.
\begin{equation}
V(r) =
\begin{cases} 
    V_{\text{LJ}}(r), & r \leq r_c \\
    0, & r > r_c
\end{cases}
\label{eq:lj}
\end{equation}
 
\noindent The Lennard Jones potential function is defined in part: When $r \leq r_c$, the potential is calculated as $4\epsilon \left[ \left(\frac{\sigma}{r}\right)^{12} - \left(\frac{\sigma}{r}\right)^6 \right]$, which captures both short-range repulsion and long-range attraction. For $r > r_c$, the potential is set to zero, reflecting the computational practice of truncating interactions beyond the cut-off to save resources. In addition, the inclusion of a cut-off distance makes the function practical for large-scale molecular systems. \\

\noindent The total internal energy $U_{\text{pair}}$ is calculated by summing the energies of the pairs of particles interacting. This is given by the following equation, where $V(r_{ij})$ is the simulated pair potential:
\begin{equation}
U_{\text{pair}} = \sum_{i=1}^{N} \sum_{j=i+1}^{N} V(r_{ij}), \quad \text{where} \quad r_{ij} \leq r_{\text{c}}.
\end{equation}

\noindent The neglected part of the Lennard-Jones potential can be approximately included by incorporating a “Long-Range Correction” (LRC), also known as “tail corrections".
This incorporates the ensemble-averaged energy contribution of the particles beyond the cut-off radius, in a manner that only depends on the density of the system and does not require pairwise distance calculations \cite{wood_monte_1957}.
The LRC is given by:
\begin{equation}
    U_{LRC} = \frac{1}{2} 4\pi \rho \int_{r_c}^{\infty} \, r^2 V (r) dr
    \label{eq:LRC}
\end{equation}
where $\rho$ is the density of the system, $r_c$ is the cut-off radius, and $V(r)$ is the pairwise energy function.\\
When $V(r) = 4\epsilon\left(\left(\frac{\sigma}{r}\right)^{12} - \left(\frac{\sigma}{r}\right)^6\right)$, this integrates to:
\begingroup
\addtolength{\jot}{1em}
\begin{align}
    U_{LRC} & = \frac{1}{2}4\pi \rho \int^\infty_{r_c} r^2 V_{LJ}(r) dr \label{eq:LRCderivation} \\ 
            & = \frac{1}{2}4\pi \rho \int^\infty_{r_c}  r^2 4\epsilon\left( \left(\frac{\sigma}{r}\right)^{12} - \left(\frac{\sigma}{r}\right)^6\right)dr \\
            %& = \frac{1}{2}4\pi \rho \int^\infty_{r_c}  4\epsilon\left(\left(\frac{\sigma^{12}}{r^{10}}\right) - \left(\frac{\sigma^6}{r^4}\right)\right)dr \\
            & = \frac{8\pi\epsilon\rho}{r_c^3}\left(\frac{\sigma^{12}}{9r_c^6} - \frac{\sigma^6}{3}\right)
\end{align}

\endgroup

\section{Formally Defining the Mathematics}\label{sec:formal}

\begin{figure}[h!]
    \centering
    \includegraphics[width=\textwidth]{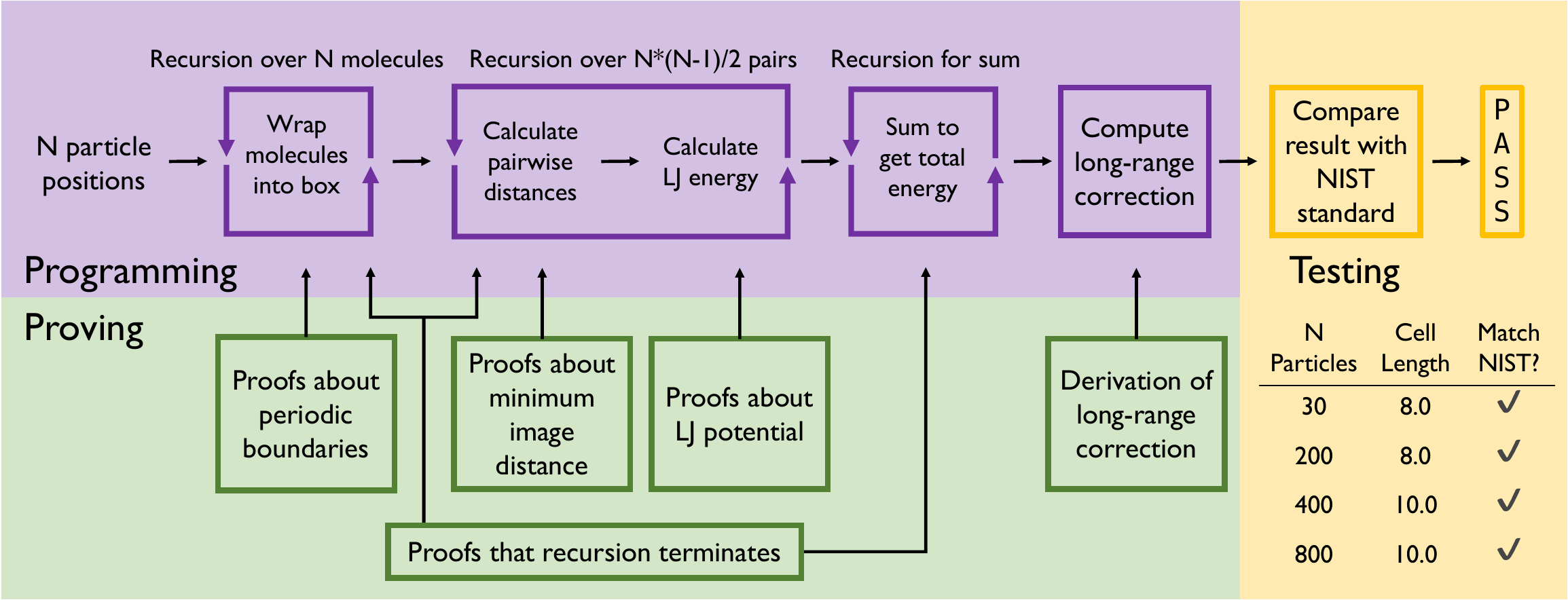}
    \caption{Workflow of the Lennard-Jones energy calculation using LeanLJ. The process involves recursive programming, formal proofs, and comparison to NIST benchmarks.}
    \label{fig:workflow}
\end{figure}

\noindent The previous section was an \emph{informal} description of these concepts; now, we turn to a \emph{formal} description, expressed as Lean code. Lean provides a structured framework to rigorously define the components of our system and prove their properties. Figure \ref{fig:workflow} illustrates our code and the means by which it is verified. In this section and the next, we describe the components of the system. We start by illustrating Lean's capabilities as a theorem prover. \\

\subsection{Introduction to Lean syntax}

\noindent Here are a few examples to illustrate the syntax of Lean 4. Lean's basic objects include types, tactics, definitions, and theorems; we do not introduce any custom types or tactics in this work, so we will focus on definitions and theorems. 
A definition has the following basic structure\footnote{This overview is inspired by the presentation in \cite{tooby_physics}.}:
\begin{code}
def name_of_object (p1 : parameter1) ... : type_of_object := the_def_of_the_object
\end{code}

\noindent A theorem (or equivalently, a lemma) has the following basic structure:
\begin{code}
theorem name_of_theorem (p1 : parameter1) ... (a1 : assumption1) ... : 
    thing_to_be_proved := by 
  proof
\end{code}

\noindent Lean's rich type system enables theorems to be stated and proved; while the user writes code, Lean effectively checks the types of the objects in the code for consistency.
Type-checking a theorem object amounts to validating whether it is true.
As the user writes the steps in a theorem's proof, Lean provides a concise overview of the current proof goal, as well as the current state of the assumptions and parameters.
This information is presented in the ``Lean Infoview' in VS Code, in what is known as a \emph{tactic state}, which is organized as follows:
\begin{code}
p1 : parameter1 
...
a1 : assumption1 
⊢ current_state_of_goal
\end{code}
To learn more about Lean, we highly recommend the textbooks ``Mechanics of Proof" by Heather Macbeth \cite{macbeth_mechanics_2022} and ``Functional Programming in Lean" by David Christiansen \cite{Christiansen_functional_programming_lean}. \\

\subsection{Lennard Jones Potential}
We can write the Lennard-Jones potential energy function in multiple ways. 
In every case, we aim to formally define Eq.~\ref{eq:lj}, using a function that takes four parameters ($\epsilon$, $\sigma$, $r$, and $r_c$) and returns the energy between a pair of particles. \\

\noindent The first version of this is \texttt{lj}. In this version, all parameters are type $\mathbb{R}$, for the Real numbers.
Lean requires this definition to be prefaced with the \texttt{noncomputable} keyword. Lean tries to generate executable bytecode for its functions, but this is not possible, in general for the Real numbers. \texttt{noncomputable} signals to Lean that this definition is only for Lean to reason about in proofs.
We intend to execute other versions of this function -- see functions \texttt{lj\_Float} and \texttt{lj\_p} for computable LJ functions in Section \ref{sec:execution}.

\begin{code}
noncomputable def lj (r r_c ε σ : ℝ) : ℝ :=
  if r ≤ r_c then
    4 * ε * ((σ / r) ^ 12 - (σ / r) ^ 6)
  else
    0
\end{code}

\noindent While \texttt{lj} may be a natural way to write Eq.~\ref{eq:lj}, alternative formulations are typically used for efficient molecular simulations.
For instance, \( r^{-3} \) can be computed first, which is then squared to obtain \( r^{-6} \), which can be squared again to obtain \( r^{-12} \).
Our function \texttt{lj\_Real} reflects this idea, using intermediate variables like \texttt{r6} and \texttt{r12}. Because \texttt{lj\_Real} is also a function of Real numbers, it is also noncomputable.

\begin{code}
noncomputable def lj_Real  (r r_c ε σ  : ℝ) : ℝ :=
  if r ≤ r_c then
    let r3 := (σ / r) ^ (3 : Nat)
    let r6 := r3 * r3
    let r12 := r6 * r6
    4 * ε * (r12 - r6)
  else
    0
\end{code}

\noindent Lean allows us to formally prove the equivalence of these two forms, as shown in the theorem \texttt{lj\_eq}, allowing us to use either representations confidently.
This capability enables not only correctness, but also flexibility in implementing the most efficient forms for simulation.
Keep in mind that we \emph{do not} address floating-point or round-off errors; this guarantee holds only for idealized functions over Real numbers, which have infinite precision.
If a more-efficient version of a function is mathematically equivalent (over Reals) to a base case, but leads to more round-off errors, that would not be detected in our formulation.
A more efficient version that is not mathematically equivalent (e.g. it invokes an approximation) would be shown to be distinct by this approach. 

\begin{code}
theorem lj_eq (r r_c ε σ : ℝ) : lj_Real r r_c ε σ = lj r r_c ε σ := by
  unfold lj_Real
  unfold lj
  simp
  ring_nf
\end{code}

\noindent The theorem \texttt{lj\_eq} formally proves that \texttt{lj\_Real r r\_c $\varepsilon$ $\sigma$ = lj r r\_c $\varepsilon$ $\sigma$}. This illustrates the syntax of Lean functions: unlike Python, which uses parentheses to denote function application (e.g., \texttt{lj\_Real(r, r\_c, epsilon, sigma)}), Lean uses simple whitespace. In the expression \texttt{lj\_Real r r\_c $\varepsilon$ $\sigma$}, each argument is applied to the function from left to right, separated by spaces. Thus, \texttt{lj\_Real r r\_c $\varepsilon$ $\sigma$} represents ``apply the function \texttt{lj\_Real} to these four arguments.'' This compact syntax is helpful in mathematical reasoning, where function application is so pervasive.\\

\noindent We can also prove various mathematical properties of our \texttt{LJ} function. 
Theorem \texttt{cutoff\_behaviour} states that for any $r > r_c$, the value of the Lennard-Jones potential is zero.
(The way to read this theorem, is ``for Real numbers $\epsilon$, $\sigma$, $r$, and $r_c$, assuming $r > r_c$, this function evaluates to zero''). 
This reflects the practice of truncating the potential beyond the cut-off distance.
\begin{code}
theorem cutoff_behaviour (ε σ r r_c : ℝ) (h : r > r_c) :
    lj_Real ε σ r r_c = 0 := by
  unfold lj_Real
  simp [if_neg (not_le_of_gt h)]   
\end{code}
\\
Theorem \texttt{ljp\_eq\_le} establishes that, in $0 < r \leq r_c$, the Lennard-Jones potential is $4\epsilon \left[ \left(\frac{\sigma}{r}\right)^{12} - \left(\frac{\sigma}{r}\right)^6 \right]$. Lean can use logical operators like $\forall$ (for all) for defining properties of functions.
\begin{code}
theorem ljp_eq_le {r_c ε σ : ℝ} : ∀ r ∈ {r | r > 0 ∧ r ≤ r_c}, 
    lj_Real r r_c ε σ = 4 * ε * ((σ / r)^12 - (σ / r)^6) := by
  intro r hr
  have h_r_le_rc : r ≤ r_c := hr.2
  unfold lj_Real
  rw [if_pos h_r_le_rc]
  ring
\end{code}
\\
We also prove the continuity of the function in this range, in Theorem \texttt{ljp\_continuous\_closed\_domain} (for brevity, we just state the theorem here; the full proof is on \href{https://github.com/ATOMSLab/LeanLJ}{GitHub}).
Continuity is essential in molecular dynamics simulations because forces are evaluated on the basis of energy gradients, and discontinuities can introduce artificial forces, destabilizing numerical integration \cite{allen_computer_2017}.
Importantly, we do not, indeed we \emph{cannot}, prove that this function is continuous for the whole domain of $r$; the \texttt{LJ} function diverges at $r=0$ and undergoes a step change at $r = r_c$. 
Researchers have implemented alternative truncation methods for the \texttt{LJ} function, such as the truncated and shifted \texttt{LJ} function or the linear force shift function, which would be continuous for all $0 < r$ \cite{frenkel_understanding_2002}. 
These properties could be formalized in Lean, but in this work, we have focused on the simple \texttt{LJ} function. 

\begin{code}
theorem lj_p_continuous_closed_domain (r_c ε σ : ℝ) :
    ContinuousOn (fun r => if r ≤ r_c then 4 * ε * (((σ / r) ^ 6) ^ 2 - (σ / r) ^ 6) else 0) {r | 0 < r ∧ r ≤ r_c} := by
\end{code}
\\

\subsection{Periodic Boundaries}

\noindent We follow the formulation in Allen and Tildesley \cite{allen_computer_2017} in defining functions for wrapping molecules according to periodic boundary conditions (PBCs) and calculating the minimum image distance. 
The periodic boundary function wraps a position from anywhere in space into the bounds of the simulation box.
This function \texttt{pbc} takes in a one-dimensional position and box length and outputs a new position (all have type $\mathbb{R}$). 

\begin{code}
noncomputable def pbc_Real (pos boxLength : ℝ) : ℝ := 
  pos - boxLength * round (pos / boxLength)
\end{code}

\noindent We formally proved that the wrapped displacement produced by the periodic boundary condition function lies within the interval $[-L/2, L/2]$ for any Real coordinate $p$ and positive box length $L$. This ensures that particles always interact with the nearest periodic image, which is a key assumption in molecular dynamics simulations. The proof was constructed in Lean by expressing the wrapped position as $L \cdot \delta$, where $\delta = \frac{p}{L} - \text{round}\left(\frac{p}{L}\right)$, and rigorously showing that $|\delta| \leq \frac{1}{2}$, hence $|\text{pbc\_Real} (p, L)| \leq \frac{L}{2}$.

\begin{code}   
theorem abs_pbc_le (p L : ℝ) (hL : 0 < L) : |pbc_Real p L| ≤ L / 2 := by
  dsimp [pbc_Real]
  let δ := (p / L) - round (p / L)
  have h_eq : p - L * round (p / L) = L * δ := by 
    rw [mul_sub]
    field_simp [hL.ne']  
  rw [h_eq, abs_mul, abs_of_pos hL]
  have hδ : |δ| ≤ 1 / 2 := abs_diff_round_le_half (p / L)
  trans L * (1 / 2)
  · exact mul_le_mul_of_nonneg_left hδ hL.le
  · field_simp
\end{code}

\subsection{Minimum Image Distance}
In defining the minimum image distance, we found it more convenient to first define the squared minimum image distance, and then take the square root of that to obtain the minimum image distance.
The box length \texttt{boxLength} and positions \texttt{posA} and \texttt{posB} are vectors in \(\mathbb{R}^N\) (with $N = 3$) where each component corresponds to a coordinate in the respective dimension.
This is specified using a vector type \texttt{Fin 3 $\rightarrow$} $\mathbb{R}$.
\footnote{Lean can handle particularly rich mathematics through its use of \emph{dependent types} – types that depend on a value. Vector is an example of this -- it is a subtype of List that depends on a value, the length of the list, which in our case, is 3. This is one way in which Lean avoids runtime errors; before the code compiles, Lean can ensure that a function taking a vector of length \texttt{N} will always receive a vector of length \texttt{N}.}
The function iterates over each of the three dimension and computes a displacement, which is adjusted using the periodic boundary function \texttt{pbc\_Real}.
The adjusted displacements are squared and summed over all dimensions. 
In our \texttt{squaredminImageDistance\_Real} function, the decide tactic is employed in each invocation of the vectors posB and posA, to prove to Lean that elements 0, 1, and 2 are in scope.

\begin{code}
noncomputable def squaredminImageDistance_Real (posA posB : Fin 3 → ℝ) (boxLength : Fin 3 → ℝ) : ℝ :=
  let dx := pbc_Real (posB (0: Fin 3) - posA (0: Fin 3)) (boxLength (0: Fin 3))
  let dy := pbc_Real (posB (1: Fin 3) - posA (1: Fin 3)) (boxLength (1: Fin 3))
  let dz := pbc_Real (posB (2: Fin 3) - posA (2: Fin 3)) (boxLength (2: Fin 3))
  dx^2 + dy^2 + dz^2 
\end{code}

\noindent We can prove a neat property of how the periodic boundaries interact with the minimum image distance – that the minimum image distance between arbitrary points in space is equivalent to the minimum image distance between those points, after being wrapped into the simulation box. This is stated in theorem \texttt{squaredminImageDistance\_theorem}, which requires an inline invoking a $\lambda$ function to iterate over the box dimensions (for brevity, the proof steps are omitted here, but available on \href{https://github.com/ATOMSLab/LeanLJ/blob/f4513e07e14e7290feccf856546fe4e5a7b7ca97/LeanLJ/MinImageDistance_PeriodicBC.lean}{GitHub}). This only holds for non-zero box lengths.

\begin{code}
theorem squaredminImageDistance_theorem (boxLength posA posB : Fin 3 → ℝ)
    (hL : ∀ i, boxLength i ≠ 0) squaredminImageDistance_Real boxLength posA posB =
    squaredminImageDistance_Real boxLength (λ i => pbc_Real (posA i) (boxLength i)) (λ i => pbc_Real (posB i) (box_length i)) := by
    ...
\end{code}

\noindent Finally, the function \texttt{minImageDistance\_Real} calls the \texttt{squaredminImageDistance} function, and takes the square root to obtain the minimum image distance. 
\begin{code}
noncomputable def minImageDistance_Real (posA posB boxLength : Fin 3 → ℝ) : ℝ :=
  (squaredminImageDistance_Real posA posB boxLength).sqrt
\end{code}

\noindent We can also prove that computed distances between particles are guaranteed to be non-negative in all applications of the minimum image convention; this can be useful when non-negativity is invoked in proofs about energy calculations.\\
\begin{code}
theorem minImageDistance_real_nonneg ( posA posB boxLength : Fin 3 → ℝ) :
    0 ≤ minImageDistance_Real  posA posB boxLength := by
  unfold minImageDistance_real
  apply Real.sqrt_nonneg
\end{code}
  
\noindent We also proved that the minimum image distance between a particle and itself is always zero (theorem \texttt{minImageDistance\_self}).
\begin{code}
theorem minImageDistance_real_self (pos boxLength : Fin 3 → ℝ) : 
    minImageDistance_Real pos pos boxLength = 0 := by
  unfold minImageDistance_Real squaredminImageDistance_Real
  have h0 : pbc_Real (pos (0: Fin 3) - pos (0: Fin 3)) (boxLength (0: Fin 3)) = 0 := by
    simp [pbc_Real, sub_self, zero_div, round_zero, mul_zero, sub_zero]
  have h1 : pbc_Real (pos (1: Fin 3) - pos (1: Fin 3)) (boxLength (1: Fin 3)) = 0 := by
    simp [pbc_Real, sub_self, zero_div, round_zero, mul_zero, sub_zero]
  have h2 : pbc_Real (pos (2: Fin 3) - pos (2: Fin 3)) (boxLength (2: Fin 3)) = 0 := by
    simp [pbc_Real, sub_self, zero_div, round_zero, mul_zero, sub_zero]
  rw [h0, h1, h2]
  simp
\end{code}

\noindent While the above formulations of \texttt{pbc\_Real} and \texttt{minImageDistance\_Real} lead to valid computations and proofs, we are somewhat dissatisfied with the semantics.
The \texttt{pbc\_Real} function operates on \emph{particle positions} (i.e. $x_i$), wrapping them inside the box from outside. 
When this function is applied in the \texttt{minImageDistance} function, it is being applied to a \emph{difference} between particle positions (i.e. $x_j - x_i$). 
Lean does not complain, because in both cases, these are just real numbers, and everything checks out, but a displacement is nonetheless not the same thing as a position.
There may be a way to make this even more rigorous, by defining a custom type for positions and restricting the \texttt{pbc\_Real} function to only operate on such a type, but we kept our approach simpler for now. \\

\subsection{Long-Range Corrections}
The long-range correction, given in Eq.~\ref{eq:LRC}, is computed using the function \texttt{U\_LRC}, which depends on $\rho$, $\epsilon$, $\sigma$, and $rc$.
\begin{code}
noncomputable def U_LRC_Real (ρ ε σ rc : ℝ) : ℝ :=
  (8 * π * ρ * ε) * ((1/9) * (σ ^ 12 / rc ^ 9) - (1/3) * (σ ^ 6 / rc ^ 3))
\end{code}
We can prove that this function follows from the integral definition of $U_{LRC}$, Eq.~\ref{eq:LRCderivation}.
The integral \texttt{$\int$ (r : $\mathbb{R}$) in Set.Ioi rc} is interpreted using measure theory, and refers to an integral over the set
\texttt{Set.Ioi rc}, which is the open interval $(r_c, \infty)$.
We state the theorem here and omit the proof for brevity, the full proof is available on \href{https://github.com/ATOMSLab/LeanLJ}{GitHub}.
\begin{code}
theorem long_range_correction_equality (hr : 0 < rc) (ρ ε σ : ℝ) :
    (2*π*ρ) * ∫ (r : ℝ) in Set.Ioi rc, 4*ε * (r^2 * (((σ / r)^12) - ((σ / r)^6))) =
    U_LRC ρ ε σ rc π := by
    ...
\end{code}
\section{Code Execution}
\label{sec:execution}

Combining formal proofs with numerical computation is central to this work. In this section, we elaborate on three aspects of programming in Lean. Subsection \ref{sec:functional} introduces the function for energy summation; in Lean, this must be recursive instead of based on traditional for loops. Subsection \ref{sec:polymorphism} highlights our approach for bridging computations and proofs using polymorphic functions. Subsection \ref{sec:monads} describes Lean's approach to input and output.

\subsection{Functional Programming}
\label{sec:functional}
Traditional molecular simulation software is implemented using imperative programming languages (like C and FORTRAN), but Lean is a functional programming language (like Haskell).
Imperative programs are about ``doing'' (following a step-by-step procedure), while functional programs are about ``being'' (defining what a function \emph{is}, which in Lean, ultimately enables proofs about it). 
Imperative programming is susceptible to ``side effects'' that are avoided in functional programming, reducing security risks and improving rigour. 
Functional programming avoids mutable data types; rather than updating (mutating) existing variables, such as assigning \texttt{x=x+1}, when new things must be computed, new variables are assigned. 
Lean 4 does support some imperative design patterns, but to get guarantees that come from proofs, writing code in a functional style is generally preferred. \\

\noindent One of the most stark differences (and most relevant for molecular simulations) between imperative and functional programming is the use of for- and while-loops.
Pairwise energy calculations typically first loop over particles $i$ from 1 to $N$, then over particles $j$ from $i + 1$ to $N$ \cite{allen_computer_2017, frenkel_understanding_2002}.
This ``double for loop'' can be expressed in Lean as follows:
\begin{code}
-- Imperative style, double for loop
def total_energy_loop (positions : List (Fin 3 → Float)) (boxLength : Fin 3 → Float) (cutoff ε σ : Float) : Float :=
  Id.run do
    let mut energy := 0.0
    for i in [0 : positions.length] do
      for j in [i+1 : positions.length] do
        let r := minImageDistance positions[i]! positions[j]! boxLength
        let e := lj_Float r cutoff ε σ
        energy := energy + e
    return energy 
\end{code}

\noindent Lean sets up for loops using the keywords \texttt{Id.run}, \texttt{do}, and \texttt{for}, and uses \texttt{mut} to denote \texttt{energy} as a mutable variable. To be clear, Lean isn't actually running imperative code; \texttt{do} notation is ``syntax sugar'' for purely functional code, which is guaranteed to be free of side effects. Lean 4's ``functional but in-place'' paradigm for memory management makes it quite efficient compared to other functional programming languages \cite{deMoura_system_description_lean}.
Lean converts the above for loops into recursive functions. %that computes the energy and adds it to a recursively-called energy calculation on a List with one fewer pairs.
In the following function we illustrate, making the recursion explicit:
\begin{code}
-- Recursive style 
def total_energy_recursive (positions : List (Fin 3 → Float))
    (boxLength : Fin 3 → Float) (cutoff ε σ : Float) : Float :=
  let numAtoms := positions.length
  let rec energy : Nat → Nat → Float → Float
    | 0, _, acc => acc
    | i+1, 0, acc => energy i (i - 1) acc
    | i+1, j+1, acc =>
      let r := minImageDistance positions[i]! positions[j]! boxLength
      let e := lj_Float r cutoff ε σ
      energy (i+1) j (acc + e)
  energy numAtoms (numAtoms - 1) 0.0  
\end{code}
\noindent Here, a recursive function \texttt{energy} is defined locally, as well an accumulation variable \texttt{acc}. 
In the central function call, \texttt{energy (i+1) j (acc + lj\_Float r r\_c $\epsilon$ $\sigma$)}, \texttt{energy} adds one \texttt{LJ} energy contribution to the value of \texttt{acc}, using particle indices $i+1$ and $j$ to obtain the distance $r$. The remaining conditions handle increments on the edge cases.

However, we found neither of these functions to be particularly amenable to proofs.
We are particularly looking for a proof that the total number of pairs in the system is $N*(N-1)/2$.
To do so, we defined a helper function that generates the pairs.
%This is similar to the function \texttt{List.sym2} in Mathlib, which returns all unordered pairs (thus, counting $i=j$); our function returns unique pairs for which $i<j$.
Our function \texttt{pairs} returns a list of pairs of atom indexes (e.g. [(0,1), (0,2), (1,2)] when $n=3$). 
\begin{code}
 -- Helper function to generate pairs
def pairs (n : Nat) : List (Nat × Nat) :=
  (List.range n).flatMap fun i =>
    (List.range' (i + 1) (n - (i + 1))).map fun j => (i, j)
\end{code}

\noindent We prove the length of this list is $N*(N-1)/2$ in \texttt{pairs\_length\_eq} in a long proof that uses mathematical induction (full proof available on \href{https://github.com/ATOMSLab/LeanLJ/blob/main/LeanLJ/Pairs_Proof.lean}{LeanLJ GitHub}). 
In the absence of test data, such a proof could provide confidence that the implementation is correct.
We anticipate that such proofs would be useful for validating energy calculations over molecules with complex exclusion rules (like excluding adjacent up to 1-4 interactions in molecules), and higher-order interactions, like 3-body potentials.
\begin{code}
theorem pairs_length_eq (n : Nat) : (pairs n).length = n * (n - 1) / 2 := by    
\end{code}

\noindent We use \texttt{pairs} in \texttt{total\_energy\_pairs}, a function that uses a simpler version of the recursive style to accomplish this summation.
This uses \texttt{foldl}, a function from functional programming that recursively applies a function to all elements of a list and accumulates the result, ``folding'' them together from the left.
\begin{code}
-- Total energy function using pairs
 def total_energy_pairs (positions : List (Fin 3 → Float))
    (boxLength : Fin 3 → Float) (cutoff ε σ : Float) : Float :=
  let n := positions.length
  let indexPairs := pairs n
  indexPairs.foldl (fun acc (i, j) => 
    let r := minImageDistance_Float positions[i]! positions[j]! boxLength
    acc + lj_Float r cutoff ε σ
  ) 0.0   
\end{code}

\noindent Lean automatically checks functions for termination, which is quite important for recursive functions, lest they get trapped in an infinite loop. 
These are verified formally using tactics that operate behind-the-scenes; only in more complicated cases, in which Lean cannot infer termination automatically, will the user be prompted to write a termination proof.
Likewise, Lean also ensures that array indexing is safe, preventing all runtime errors associated with accessing out-of-bounds indices.
In developing this code, we found execution and comparing to the NIST tests was valuable for developing the logic of the loops and the recursion, as it enabled quick feedback.
Our code also incorporates tail recursion to facilitate efficient execution \cite{Christiansen_functional_programming_lean}.

\subsection{Polymorphism}
\label{sec:polymorphism}

\begin{figure}
    \centering
    \includegraphics[width=0.4\textwidth]{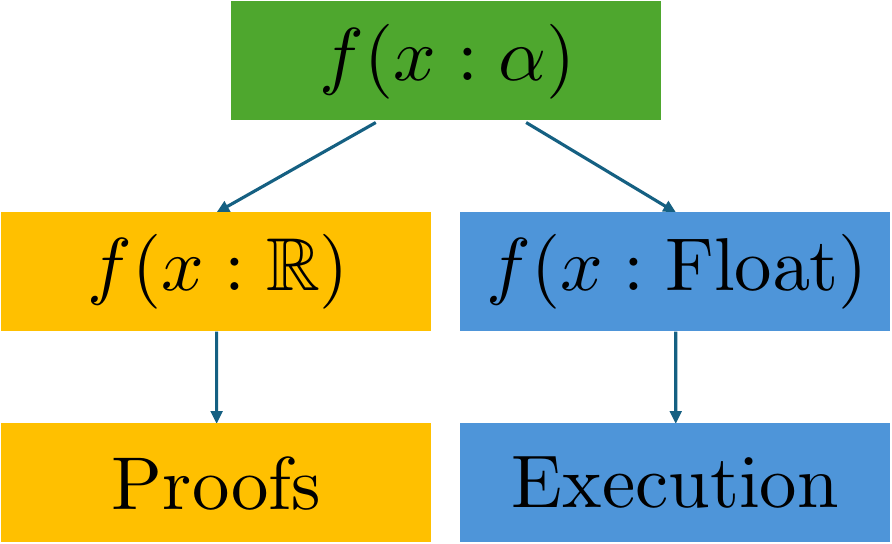}
    \caption{How polymorphic functions link proofs (over idealized Real numbers) with execution (over floating point numbers). The polymorphic function $f$ is defined for $x$ with generic type $\alpha$; proofs about $f$ can be written when $x$ is Real, and computations with $f$ can be executed when $x$ is a float.}
    \label{fig:polyscheme}
\end{figure}

In Lean, we can define functions specifically for Real numbers (\(\mathbb{R}\)), which allows us to prove mathematical properties, or for floating-point numbers (\texttt{Float}), which enables efficient numerical computation. However, these separate implementations create a trade-off: the \texttt{Real} version is \textit{non-computable}, meaning it cannot be executed in actual simulations, while the \texttt{Float} version is not suitable for formal proofs, as floating-point arithmetic lacks the necessary mathematical structure (in the typical standard for floating point addition, IEEE 754, 0.1 + 0.2 $\neq$ 0.3).
\noindent To bridge this gap, \textit{polymorphic functions} are used, allowing the same definition to work for multiple types (Fig.~\ref{fig:polyscheme}).
By introducing a generic type \(\alpha\) that can subsume both \texttt{Real} and \texttt{Float}, we ensure that our function can operate on both Reals (\(\mathbb{R}\)) for proofs and floats (\texttt{Float}) for computations. \\

\begin{figure}
    \centering
    \includegraphics[width=0.8\textwidth]{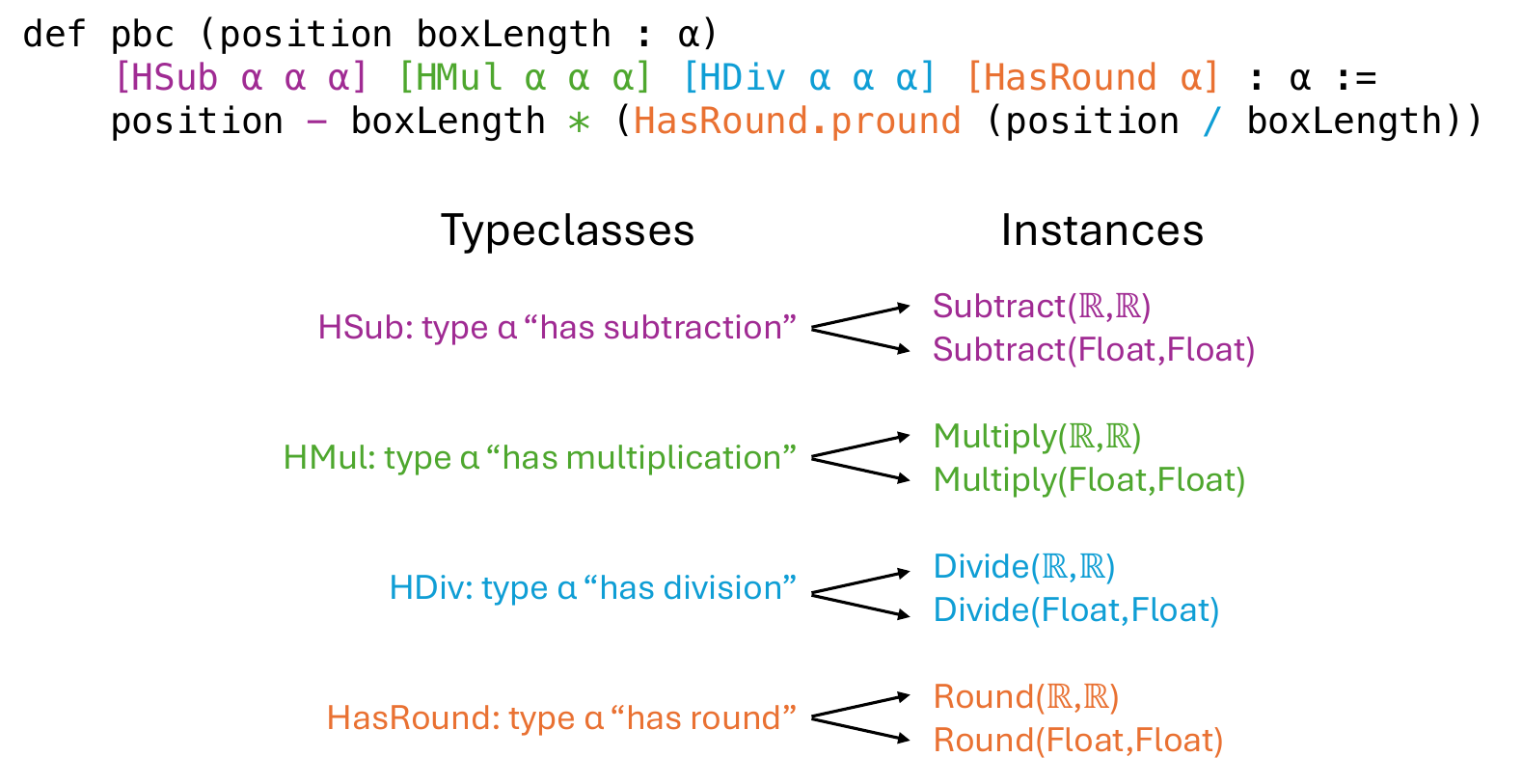}
    \caption{Explanation of the polymorphic \texttt{pbc} function. The function is defined over a generic type $\alpha$, and the required operations---subtraction, multiplication, division, and rounding---are expressed through type classes: \texttt{HSub}, \texttt{HMul}, \texttt{HDiv}, and \texttt{HasRound}. Each type class specifies that the type $\alpha$ must support a given operation. For example, \texttt{HSub $\alpha$ $\alpha$ $\alpha$} means $\alpha$ must support subtraction with two $\alpha$ inputs returning an $\alpha$ result. Concrete instances, such as \texttt{Float} and $\mathbb{R}$, implement these type classes to enable polymorphic behaviour. This allows \texttt{pbc} to work with different numeric types, as long as they satisfy the required operations.}
    \label{fig:polymorphic-pbc}
\end{figure}

\noindent To illustrate, consider the function \texttt{pbc} (Fig.~\ref{fig:polymorphic-pbc}), which wraps a particle's position into the simulation box using periodic boundary conditions.
Section \ref{sec:formal} showed \texttt{pbc\_Real}, which operates on position and box length with type $\mathbb{R}$; \texttt{pbc} operates on position and box length with generic type $\alpha$. 
We tell Lean more about what $\alpha$ can be, using \emph{type classes} and \emph{instances}. Specially, \texttt{pbc} is defined for any type $\alpha$ that ``knows how to'' subtract, multiply, divide, and round. These capabilities are provided through the type classes \texttt{HSub}, \texttt{HMul}, \texttt{HDiv}, and \texttt{HasRound}.
For example, the type class \texttt{HSub [$\alpha$ $\alpha$ $\alpha$]} requires that there exists a definition of subtraction between two members of $\alpha$ that would output a third member of $\alpha$.
\texttt{HSub}, \texttt{HMul}, and \texttt{HDiv} are all defined in Mathlib; for rounding, we defined a custom type class, since Mathlib did not already define that connection.
This approach allows our definition of \texttt{pbc} to  be used in two very different ways: with Real numbers for formal proofs, and with floating-point numbers for actual simulations. \\

\noindent Prefacing each function with a long list of instances is rather cumbersome. 
We can group all of these instances together into a single type class, \texttt{RealLike}, named thus because it is ``like'' Real numbers.
\texttt{RealLike} is ultimately compatible with Floats, but it doesn't have all the properties of the Real numbers (after all, it cannot use functions we did not attach to the type class). The \texttt{RealLike} type class captures the essential features of types that behave like real numbers, making it easier to write numeric code that works across different types such as \texttt{Float} or \texttt{Int}. It includes basic operations like addition, subtraction, multiplication, division, negation, and exponentiation with natural numbers that appear in our equations (1, 3, 9, etc.). It also provides comparisons (\(\le\), \(<\)) that return \texttt{Bool}, along with constants like zero and one, and ways to convert from natural and integer literals. In addition, \texttt{RealLike} extends two smaller type classes---\texttt{HasSqrt} for square roots (used in distance calculations) and \texttt{HasRound} for rounding (used in periodic boundaries) so that any type marked as \texttt{RealLike} is also expected to support those operations. This setup allows us to write clean, reusable, and type-safe numeric functions without tying them to one specific number type.\\

\noindent We note that this type class approach admits a small possibility of error that Lean will not catch. The RealLike type class manually links certain Real-typed and Float-typed functions, and mistakes in RealLike won’t be flagged by Lean. For example, a subtle error in which one rounding function rounded 0.5 \emph{down} while another rounded 0.5 \emph{up} would not be detected by Lean. In fact, egregious errors linking a square root and a cube root are also technically possible, so human oversight remains necessary with this approach. Essentially, the RealLike type class defines which Real and Float functions are \emph{semantically} equivalent, and the human is responsible for ensuring correct semantics. The Lean community is developing more systematic solutions, such as the ComputableReal project led by Alex Meiburg, which provides a more principled, computable representation of real numbers that integrates cleanly with existing  numeric type classes \cite{alex_meiburg_computablereal_2025}.

\begin{code}
class RealLike (α : Type) extends HasRound α, HasSqrt α where
  add : α → α → α
  sub : α → α → α
  mul : α → α → α
  div : α → α → α
  neg : α → α
  pow : α → Nat → α
  le : α → α → Bool
  lt : α → α → Bool
  zero : α
  one : α
  ofNat : Nat → α
  ofInt : Int → α
\end{code}

\noindent For some more examples, we provide the implementations for the Lennard-Jones potential in three forms: the \textit{polymorphic} version ($\alpha$), the \textit{Real number} version (\(\mathbb{R}\)), and the \textit{floating-point} version (\texttt{Float}).

\begin{code}
-- Polymorphic version: Works for both ℝ and Float
def lj_p {α : Type} [RealLike α] (r r_c ε σ : α) : α :=
    if r ≤ r_c then
      let r3 := (σ / r) ^ (3 : Nat)
      let r6 := r3 * r3
      let r12 := r6 * r6
      4 * ε * (r12 - r6)
    else
      0
\end{code}
\begin{code}
-- Real number version: Allows formal proofs but cannot compute
noncomputable def lj_Real  (r r_c ε σ  : ℝ) : ℝ :=
    if r ≤ r_c then
      let r3 := (σ / r) ^ (3 : Nat)
      let r6 := r3 * r3
      let r12 := r6 * r6
      4 * ε * (r12 - r6)
    else
      0 
\end{code}
\begin{code}
-- Floating-point version: Can compute but lacks proof capabilities
def lj_Float (r r_c ε σ : Float) : Float :=
    if r ≤ r_c then
      let r3 := (σ / r) ^ (3 : Nat)
      let r6 := r3 * r3
      let r12 := r6 * r6
      4 * ε * (r12 - r6)
    else
      0
\end{code}

\noindent Ultimately, we use \texttt{RealLike} to define polymorphic versions of \emph{all} executable functions in the overall execution flow (Fig.~\ref{fig:workflow}) and connect them to their Real counterparts.
The polymorphic version (with type \texttt{RealLike}) of each function is that which is ultimately executed; the Real version or the polymorphic version is used in the proofs.

\subsection{Input and Output in Lean}
\label{sec:monads}
Most of Lean is developed in terms of pure functions, whose behaviour can be guaranteed because the argument types limit the domain of the function inputs. By chaining pure functions with pure functions through-and-through, Lean guarantees there are no side effects. 
But input/output (IO) operations cannot have the same guarantees.
For instance, if one writes a molecular configuration file to disk, then reads it back in, one cannot guarantee that some other process modified it in the meantime. \\

\begin{figure}[h]
    \centering
    \includegraphics[width=0.8\textwidth]{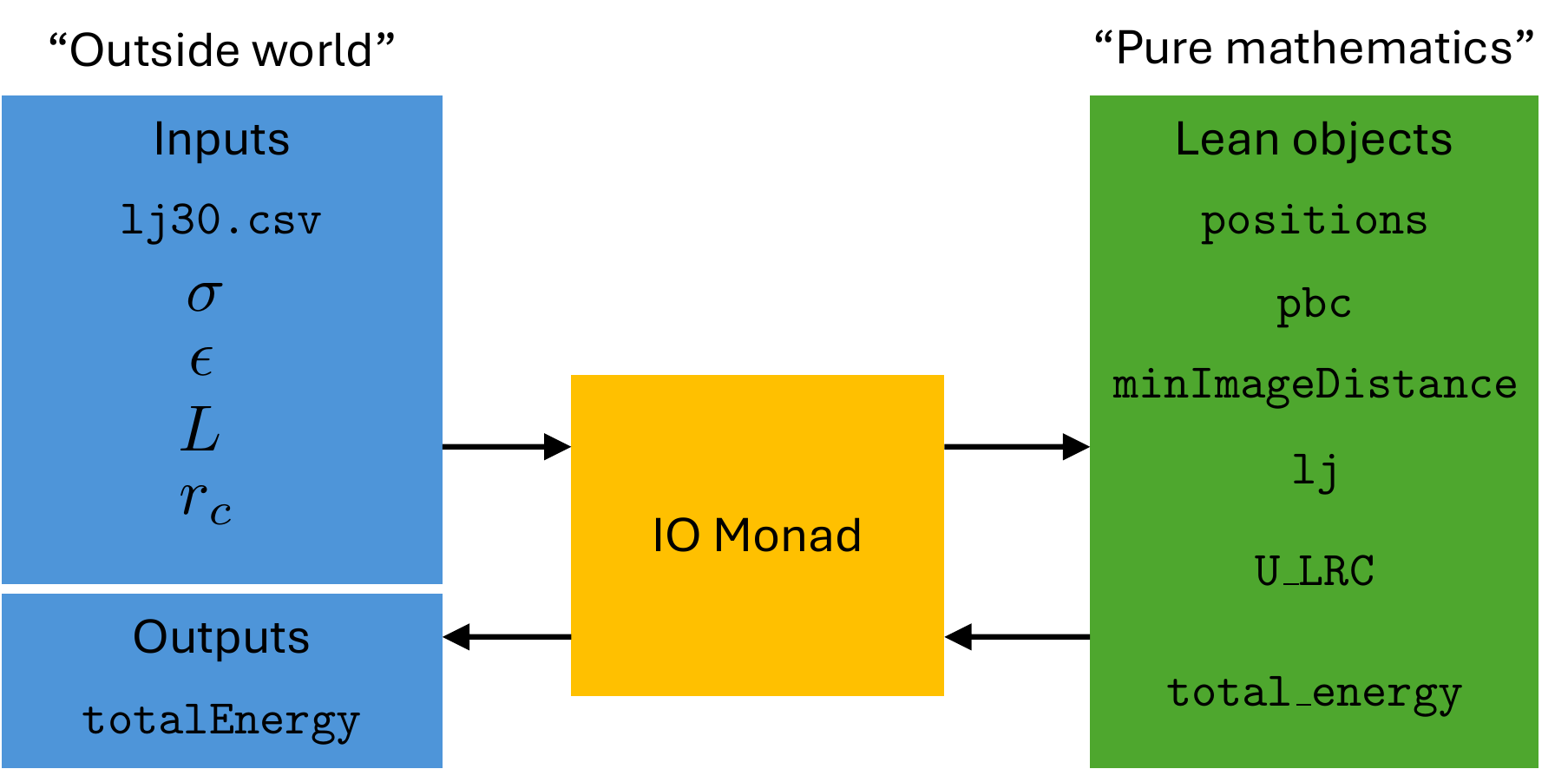}
    \caption{The IO Monad as a bridge that links the verified, pure functions in Lean with the messy real world, where data and simulation inputs reside. 
    The CSV parser uses the IO Monad to read the particle coordinates from lj30.csv into the Lean object \texttt{positions}.}
    \label{fig:io_monad}
\end{figure}

\noindent But to be a useful programming language, Lean must nonetheless have IO.
Lean separates this cleanly from its pure functions and math libraries, implementing it in the \texttt{IO} monad.
This essentially serves as a bridge between the messy, ``outside'' world and the safe, pure functions inside Lean (Figure \ref{fig:io_monad}). \\

\noindent Monads are used to handle many kinds of computation patterns in a clean and consistent way, such as optional values, errors, and non-determinism. For example, the \texttt{Option} monad handles missing values, the \texttt{Except} monad deals with errors without crashing, and the \texttt{List} monad allows multiple possible results from a single computation. They are central in functional programming, but are encountered less often in imperative languages; the interested reader can learn more here \cite{Christiansen_functional_programming_lean}. \\

\noindent To import the configuration files from the NIST SRSW, we first saved them as comma-separated values (CSV) files.
We adapted the CSV reader and used it to parse each configuration.
In addition, users are asked to manually enter simulation parameters such as the cut-off radius, $\sigma$, $\varepsilon$, and the length of the box through the terminal.
These user inputs and file reads are examples of interaction with the ``outside world,'' and are handled explicitly in Lean using the IO monad.
This makes it clear which parts of the program remain exposed to sources of error -- our setup does \emph{not} provide guarantees against sources of error on the ``outside'' of IO; if an incorrect value for $\sigma$ were input, Lean would not catch it.
This would be a form of semantic error (Table \ref{tab:errors}) that our current implementation does not avoid. \\

\noindent Proofs in Lean only provide guarantees about pure functions; errors in the I/O layer cannot be validated in this manner.
This is why we advocate for both proofs and tests (Fig.~\ref{fig:correctness-approaches}). 
For instance, while developing this application, our first approach for reading the configuration failed to read all atoms, leading to incorrect energy calculations. 
Our proofs do not catch bugs like these, but the tests do.

\section{Results}
\label{sec:results}

To evaluate our implementation, we compare the pairwise interaction energy ($U_{\text{pair}}$) and long-range correction (LRC) values computed using our Lean code with the NIST Standard Reference Simulation Website (SRSW) benchmark values \cite{shen_nist_2017} for \texttt{LJ} particles in a cubic box (Table~\ref{tab:nist-comparison}). The results show exact agreement for all four systems, within the number of digits provided by NIST.

%\clearpage
\begin{table}
\centering
\caption{Comparison of LRC and $U_{\text{pair}}$ energy calculations from NIST SRSW  \cite{shen_nist_2017} and LeanLJ for various particle counts. Energies are reported in scientific notation (reduced units), with one more digit than NIST.}
\label{tab:nist-comparison}
\begin{tabular}{|c|c|c|c|c|}
\hline
\textbf{Particles} & \textbf{$U_{\text{pair}}$ (Lean)} & \textbf{$U_{\text{pair}}$ (NIST)} & \textbf{LRC (Lean)}  & \textbf{LRC (NIST)} \\
\hline
30  &  \(-1.67903\mathrm{E}{+01}\) & \(-1.6790\mathrm{E}{+01}\)  & \(-5.45166\mathrm{E}{-01}\) &
\(-5.4517\mathrm{E}{-01}\) \\

200 & \(-6.90004\mathrm{E}{+02}\) & \(-6.9000\mathrm{E}{+02}\) & \(-2.42296\mathrm{E}{+01}\) & \(-2.42296\mathrm{E}{+01}\)   \\

400 & \(-1.14666\mathrm{E}{+03}\) & \(-1.1467\mathrm{E}{+03}\) & \(-4.96222\mathrm{E}{+01}\) &
\(-4.9622\mathrm{E}{+01}\)  \\

800 & \(-4.35154\mathrm{E}{+03}\) & \(-4.3515\mathrm{E}{+03}\)  & \(-1.98488\mathrm{E}{+02}\) &
\(-1.9849\mathrm{E}{+02}\) \\
\hline
\end{tabular}
\end{table}

\section{Discussion and Outlook}
In this study, we developed pairwise energy calculations in Lean and compared our results with the values provided by the NIST SRSW benchmark. Our calculations agree to machine precision with the NIST reference values. 
To be clear, our confidence in our system does \emph{not} stem from its agreement with the NIST benchmark, rather from the theorems we have proved in Lean that certify that the functions in LeanLJ have those specified mathematical properties. 
For instance, the \texttt{pbc} function guarantees that all wrapped particles lie in the interval $[-L/2, L/2]$, and the derivation of the function computing long-range corrections is validated mathematically.
We assert that LeanLJ is a more reliable benchmark than the NIST SRSW, at least for the components of the benchmark we have addressed.
LeanLJ could be validated even further by adding to the list of proved theorems about current functions.

\noindent We consider it helpful to reflect on the remaining sources of uncertainty in our code -- considering what we have verified, what could still be wrong? 
First, we are trusting in the axioms of mathematics, as expressed in Lean's core; errors here might compromise Mathlib, on which we depend. 
Second, our approach to polymorphism exposes us to mistakes in our \texttt{RealLike} type class, as we described in Section~\ref{sec:polymorphism}; Lean does not check to ensure that \texttt{RealLike} links the correct Float- and Real-type functions.
Third, when we developed both polymorphic \emph{and} floating-point versions of the same function for illustration purposes; we sometimes mistakenly called the floating-point version in execution instead of the polymorphic version. 
In this case, proofs for that function are not technically connected to the execution -- since Lean does not require our polymorphic code design, it does not flag such mistakes (these could be eliminated by not defining Float and RealLike versions of the same functions).
Fourth, we are still exposed to errors in input/output (Section \ref{sec:monads}), and in defining system-specific parameters, such as the force field parameters; these are mitigated by the testing, but do not prevent a user from inputting incorrect parameters for calculations outside the scope of the NIST benchmarks.
A fifth source would be vulnerabilities in the broader operating system in which the code is executed.
Nonetheless, traditional molecular simulation have far more possibilities for errors, such that many of these concerns are not considered in typical conversations about software correctness. \\

\noindent More broadly, this work demonstrates how Lean can provide a new paradigm for computational molecular simulations, where the results and the entire computational process are provably correct.
Logical steps to build on this framework include implementing support for triclinic simulation boxes, Ewald summation for Coulomb interactions, neighbour lists to improve computational efficiency, and of course, integrating Newtons equations of motion to evolve particle trajectories.
Some of these are matters of implementation (triclinic cells), but others will involve grappling yet-unresolved questions of how to handle various approximations in a formal environment, such as how to precisely describe the conditions under which neighbour lists can be trusted.\\

\noindent Nonetheless, we believe that Lean can be used (in principle) to implement molecular simulation software with all the features of the established packages used in daily practice.
The user interface would not need to be more complex; the proofs and theorems would all be handled in the back-end.
Furthermore, this could be integrated with software frameworks like MoSDeF \cite{summers_mosdef_2020} \cite{craven_achieving_2025}, that generate input files for multiple simulation engines. 
This would help address reproducibility and rigour at the ``outside world'' layer (Fig.~\ref{fig:io_monad}) in which the force field parameters, molecular specifications, and configuration are defined.
This would also enable rigorous evaluation of traditional software using Lean software as a benchmark; MoSDeF could create inputs for both verified Lean software as well as for LAMMPS, and deviations between the respective outputs may signal a bug in LAMMPS.
\\

\noindent In our previous work, we showed Lean's broader utility for formalizing derivations in science as math proofs  \cite{bobbin_formalizing_2024}, digitizing key results in absorption theory, thermodynamics, and kinematics. Joseph Tooby-Smith is also developing derivations in the high-energy physics field \cite{tooby_physics, tooby-smith_formalization_2024}. 
These early works showcase Lean's rigour and versatility for building a library (or libraries) of formally-verified results in diverse areas of science, facilitating rigorous verification of scientific ideas in different disciplines. In the long term, we envision Lean being used to formalize not just interaction potentials or particle dynamics, but the very foundations of statistical mechanics itself.
The mathematics of ensemble theory, such as definitions of the microcanonical (NVE) and canonical (NVT) ensembles and proofs about their properties, could be stated precisely, and then directly linked to formally verified simulation code.
For instance, we imagine proving that a molecular dynamics simulation routine satisfies the conservation laws of the NVE ensemble, in the limit of infinitesimal time steps, and verifying the detailed balance condition of Monte Carlo moves. %This level of rigour is literally impossible to achieve with conventional programming languages, which lack the capability to express and verify mathematical theorems.\\ %Consequently, after a rigorous foundation is established, novel algorithms and simulation methods (crafted by humans and/or AI) could be built on top, along with correctness guarantees. % While this is still far from reality, this work represents a foundational step in that direction starting with the simplest of systems and gradually scaling up toward the grand challenge of provably correct statistical mechanics.\\

\noindent LeanLJ demonstrates how \emph{executable} scientific computing software can be tied to such proofs, using polymorphic functions. We believe this approach is quite general for reasoning about idealized Real-valued functions in scientific \emph{theories}, while linking these to floating-point executions in scientific computing software.
Certigrad's \cite{selsam_developing_2017} approach is also worth considering; this verifies the high-level mathematics in Lean, and then links high-level functions to unverified, but efficient, linear algebra libraries written in C.
Compared to our approach, Certigrad's ``bridge'' between verified math and executable math consequently happens at a higher level; our polymorphic functions build this bridge at the level of individual math operators (e.g. addition, division) and constants (e.g. $\pi$). \\

\noindent Scientific computing benchmarks are typically based on human oversight and software best practices \cite{Thompson_Molecular_reproducibility}; formal Lean verification offers an even more rigorous alternative, allowing rigorous mathematical proofs that the implemented software is correct. This shift from empirical validation to formal proof introduces a new level of confidence in molecular simulations, setting the stage for more reliable and mathematically sound scientific computing.

\section*{Conclusion}
This work demonstrates how formal methods can complement molecular simulations by providing proven guarantees about the properties of the molecular simulation system.
While still in its early stages, this approach opens a path toward mathematically-verified simulations.

\section*{Acknowledgements}
The authors thank the members of the Lean Zulip for helpful discussions, especially Tomáš Skřivan for his insights into polymorphic functions. This material is based on work supported by the National Science Foundation (NSF) CAREER Award \#2236769.

\section*{Declaration of Interest}
The authors declare no competing financial or personal interests that could influence the work reported in this paper.

\section*{Data Availability}
All code, proofs, and benchmark files are on the \href{https://github.com/ATOMSLab/LeanLJ}{ATOMS Lab Github}. 

\section*{ORCID}

\noindent
Ejike D. Ugwuanyi \href{https://orcid.org/0009-0006-4335-5428}{\texttt{https://orcid.org/0009-0006-4335-5428}} \\
Colin T. Jones \href{https://orcid.org/0009-0007-9013-8036}{\texttt{https://orcid.org/0009-0007-9013-8036}} \\
John Velkey \href{https://orcid.org/0000-0001-5156-7451}{\texttt{https://orcid.org/0000-0001-5156-7451}} \\
Tyler R. Josephson \href{https://orcid.org/0000-0002-0100-0227}{\texttt{https://orcid.org/0000-0002-0100-0227}}

\bibliographystyle{unsrturl}
\begin{spacing}{0.5}
\bibliography{references.bib}
\end{spacing}

\clearpage

%%%%%%%%%%%%%%
\end{document}